# *Biophysical analysis of electric current mediated nucleoprotein inactivation process.*

*José E. González[1*], Robersy Sánchez[1,2] and Aminael Sánchez[2,3]*

[1]Institute of Tropical Crops Research, INIVIT. P.O. Box 6, Santo Domingo, Cuba.
[2]Bioinformatics Group, Center of Studies on Informatics, Central University of 'Las Villas', 54830 Cuba.
[3]Institute of Plants Biotechnology, Central University of 'Las Villas', 54830 Cuba.

**ABSTRACT**

**Motivation:**

Its well known the use of electric current in the cleaning of viral diseases in plants instead a deep knowledge of this phenomenas theoretical basis is not yet available. The description of the real causes of nucleoprotein inactivation, will contribute to the optimization of further experiments in order to obtain more and more efficient cleaning methodologies for starting material supplies of vegetal species micropropagation process. The dissipated energy as heat will depend on the specific treatment applied and on the physical properties of the vegetal tissue and viral nucleoprotein.

**Results:**

A hyperbolic dependence between the absorbance output values, obtained during a diagnostic micro-ELISA experiment, and the electrical power (in watts) applied to each explant was found. The former, demonstrate the cleaning process nature is essentially the effect over viral nucleoprotein denaturation, by means of the heat, to which they were exposed in the thermal bath that the vegetal tissue constitutes. Our results are consistent with mathematical developed from theoretical frame of harmonic oscillators in which molecular machine (viral nucleoproteins) could be redefined.

**Contact:** E-mail: josee@inivit.co.cu, gefrain00@yahoo.es

## INTRODUCTION

**Electric current mediated cleaning of viral diseases in plants.**

Traditional techniques applied to cleaning of viral diseases in plants, i.e. Meristems culture, thermotherapy and chemotherapy fail to produce enough quantities of clean material in most species. Alternative procedures using electric current treatments have become in an efficient tool to overcome this problem.

Black (1971) demonstrated a relation between growths stimulation in tomato plants that were treated with low current densities (3-15 μA/plants) during 4, 5, 12 and 24 hours and the ion concentrations detected. Blanchard (1974) applying 1 - 4 amperes DC pulsed to 6500 volt/hours, after 2 - 3 days, obtained clean tissues without microflora presence. Those results settled down the basis for the electrotherapy concept.

Wagele (1978) patented his experience to use electrical current in obtaining grown bacterial cultures and tissues from various sources.

Quacquerelli *et al.*, (1980) applied electric current to *Cactanucia* tree stakes showing intense mosaic symptoms caused by virus, proving that treatments of 500 V/5-10 min. lead up to 90% of cleaned plants. Electrotherapy was then recommended in cleaning of Cucumber Mosaic Virus, Arabia Mosaic Virus, Grapevine Franleaf Virus, Chicory Yellow Motle and Tobaco Mosaic Virus.

Based in these result, Hernández *et al.* (1997b) built a device to apply electric current in the cleaning of a viral complex in garlic (*Allium sativum* L.). They obtained an efficiency of 53-100% cleaned plant using 10-30 V treatments. Other applications of the same technology have been reported in sugar cane (*Saccharum* sp hybrid), potato (*Solanum tuberosum*) (Bernal, 1997), aroids (*Xanthosoma sagitifolium*) (Igarza et al. 2001) and banana plants (*Musa ssp*) (Hernández et al. 2002) for VMCA, PLRV, DMV and BSV respectively.

Nevertheless, the lackness of homogeneity in the test (due to differences in explant electrical resistance) leads to differences of the electrical power that circulates through each of them, when fixed voltages or electrical currents are applied.

**Molecular machines. Harmonics molecular oscillators.**

Its known that molecular structures, proteic or nucleoproteic could be redefined as molecular machines. In other words, a molecular machine is just a simple macromolecule o a macromolecular complex. A molecular machine executes specific functions in living systems, they are isothermals. It means they are forced to work in a narrow range of temperatures in the thermal bath where they exist. Nevertheless, they can use the primary energy from surrounds to change its conformation into more flexible one. That is, essentially, a controlled way of denaturation what results in a loss of its biological function. Beyond the primary energy, any excess of energy is quickly dissipated leading the molecule confinement to the previous state at physiological temperature. A specific macromolecule receives a quantum of energy that put it from a basal state to an excited state. A further specific action always occurs dissipating the absorbed energy which is coupled to the realization of a specific function, evolutionarily advantageous for the organism who synthesized the macromolecule.

Molecular machines could be stimulated by any energy source besides primary energy sources. That includes not only photons and ATP but also thermal movement like the restriction enzyme Eco*RI* isolated from its DNA binding site. Eco*RI* uses thermal fluctuations, originated by microscopical heat, as energy source to reach the DNA in a specific manner (Schneider, 1991a y 1991b).

**MATERIALS AND METHODS.**

Plants of the genotype Pacala Duclos (*Dioscorea alata* L.) serologically positives to potyviruses were used. Nodal segments (25 in total), treated with 15 volt DC for 5 minutes, were used as described by González 2005.

**Determination of the relation between Electrical Power and Absorbance Value of in the micro-ELISA test**

The electric current was measured in each of the explants. Then, the electrical power (energy vanished in form of heat) was calculated according to the equation:

P = I x V.

where:

P is Electrical power (Joule/second)

I is Current Intensity (Ampere)

V is Voltage (watts)

**Theoretical modeling of the effect of the electrical power on viral nucleoproteins.**

The Molecular Machines inside the cells can be treated, or they behave, as harmonic oscillators in thermal bathroom. If the viral particles behave as oscillators, then we should hope the population's of viral particles half energy is proportional to the magnitude RT, where R is the gases constant y T is the absolute temperature, because RT is the media energy, for particle mole, for a oscillator in balance in a thermal bath (Schneider, 1991a y b). In particular, we assume that the viral particles behave as quantum oscillators, that is to say, they can take alone discreet securities of energy:

$\varepsilon_n$ = nx$\varepsilon_o$, where n is any integer and $\varepsilon_o$ is the primary energy (joule/mole) of the particles in the thermal bath from which the virus begin to inactivate. Starting from this reasoning a theoretical model is developed to find the dependence between the Absorbance and the electrical power in each explant.

**Lineal regression of the experimental data. Fit to theoretical model.**

Using a transform dependent variable x=1/P, two linearization models were proved: a model with intercept y=bx+c+error, and a model without intercept y=bx+error (Darlington, 1990).

## RESULTS

**Determination of the relation between electrical power and the absorbance value in the micro-ELISA assay.**

Table 1 shows the measure and calculated parameters during the electrotherapy experiment. Of 25 plants included, 21 (84%) regenerated into *in vitro* plants. The efficiency reached in the assay was 64%. The differences among measured current values (difference among electrical power calculated) are due to differences of electric resistance in vegetal explants that close the electric circuits. The lackness of homogeneity in age, cellular constitution, thickness, length and hydric potential leads to a different electric resistance in each case.

The electrical power is a measure of the energy dissipated in way of heat in an electric circuit (in each explant). In a qualitative analysis the smaller the absorbance value the higher the value of electrical power applied and viceverse. This relation means that, in a practical range, the highest electrical power applied leads to reduction of the amount of viral particles detected by ELISA reaction.

Table 1 Absorbance, electric current, resistance and power values determinated for each explant.

| Explant | Current (mA) | Resistance ($\Omega$) | Power (w) | Abs. (550 nm) | Micro-ELISA (c.o$^*$ 0.21) |
|---|---|---|---|---|---|
| 1  | 54  | 277.7  | 0.81 | 0.15 | Negative |
| 2  | 101 | 148.5  | 1.52 | 0.14 | Negative |
| 3  | 80  | 187.5  | 1.21 | 0.17 | Negative |
| 4  | 36  | 416.6  | 0.54 | 0.13 | Negative |
| 5  | 61  | 245.9  | 0.92 | 0.10 | Negative |
| 6  | 120 | 125.0  | 1.81 | -    | N.R$^\dagger$ |
| 7  | 94  | 159.5  | 1.42 | 0.11 | Negative |
| 8  | 15  | 1000.0 | 0.23 | 0.58 | Positive |
| 9  | 65  | 230.7  | 0.98 | 0.15 | Negative |
| 10 | 21  | 714.2  | 0.32 | 0.38 | Positive |
| 11 | 82  | 182.9  | 1.23 | 0.08 | Negative |
| 12 | 108 | 138.8  | 1.63 | 0.09 | Negative |
| 13 | 116 | 129.3  | 1.75 | 0.13 | Negative |
| 14 | 103 | 145.6  | 1.55 | -    | N.R |
| 15 | 18  | 833.3  | 0.28 | 0.57 | Positive |
| 16 | 58  | 258.6  | 0.87 | 0.14 | Negative |
| 17 | 50  | 300.0  | 0.75 | 0.13 | Negative |
| 18 | 120 | 125.0  | 1.80 | 0.09 | Negative |
| 19 | 166 | 129.3  | 1.74 | -    | N.R |
| 20 | 126 | 119.0  | 1.90 | -    | N.R |
| 21 | 79  | 189.8  | 1.19 | 0.05 | Negative |

| 22 | 98 | 153.0 | 1.48 | 0.06 | Negative |
| 23 | 90 | 166.0 | 1.35 | 0.12 | Negative |
| 24 | 23 | 652.1 | 0.35 | 0.52 | Positive |
| 25 | 24 | 625.0 | 0.37 | 0.39 | Positive |

*c.o = Cut off output value from ELISA experiment; †N.R = no regenerated *in vitro* plants were obtained

**Theoretical modeling of the effect of the electrical power on viral nucleoproteins.**

The probability of finding viral particles inactivated in the state of energy $\varepsilon_n$ at temperature T is given by the Boltzmann distribution:

$$p_n = \frac{e^{-\frac{n\varepsilon_o}{RT}}}{\sum_n e^{-\frac{n\varepsilon_o}{RT}}}$$

If we assumed that the number of moles of inactivated particles in a system with energy $\varepsilon_n$ is $N_n = n\,N_o$, where $N_o$ is the number of moles of inactivated particles in the same system with energy $\varepsilon_o$ defined at physiological temperature from which viral particles are inactivated, then the expected number of inactivated viral particles at temperature T is given by the expression:

$$\langle N \rangle_T = \sum_n N_n p_n = N_o \frac{\sum_n n e^{-\frac{n\varepsilon_o}{RT}}}{\sum_n e^{-\frac{n\varepsilon_o}{RT}}}$$

Where R is the universal constant of gases and T is the Absolute Temperature (°K). For sufficiently large values of *n* we can approximate the previous expression. Doing $\lambda = \varepsilon_o/RT$, we have:

$$\langle N \rangle_T = -N_o \frac{\frac{d}{d\lambda}\left(\sum_n ne^{-n\lambda}\right)}{\sum_n e^{-n\lambda}} = -N_o \frac{d}{d\lambda}\left(Ln\sum_n ne^{-n\lambda}\right)$$

for sufficiently large *n*, the former tends to:

$$\sum_n ne^{-n\lambda} \approx \frac{1}{1-e^{-\lambda}}$$

$$-Ln\sum_n ne^{-n\lambda} \approx Ln(1-e^{-\lambda})$$

$$\frac{d}{d\lambda}Ln(1-e^{-\lambda}) = \frac{e^{-\lambda}}{1-e^{-\lambda}}$$

If we consider, in addition, that: $e^\lambda \approx 1+\lambda$, then replacing the corresponding expressions, finally we obtain:

$$\langle N \rangle_T = \frac{N_0}{e^\lambda - 1} = \frac{N_0}{\lambda}, \text{ that is}$$

$$\langle N \rangle_T = \frac{N_0}{\varepsilon_0} RT$$

So that, the number of inactivated viral particles increases when the temperature inside the cells of the tissue submitted to treatment increases. It is evident that the observed absorbance measures (A) in the tissue samples analyzed by immunoassay, must be inversely proportional to the average number of inactivated particles, since a linear relation exists, in the range of optical density

of the values of Table 1, between the absorbance and the amount of particles (Lambert-Beer law):

$$A = \frac{c}{\langle N \rangle_T}$$

In other words, the higher the value of inactivated viral particles the lower the value of absorbance observed. The constant of proportionality $c$ must be based on the physicochemical properties of the used tissue such as its thermal conductivity, electrical conductivity and of the used method of measurement. Finally we have:

$$A = \frac{c\varepsilon_0}{N_0} \frac{1}{RT} = A_0 \frac{1}{RT} \quad (1)$$

The inactivation by denaturation caused by the increase of viral particles energy, what does not allow particles to be recognized by the antibody mediated signaling complex in the ELISA test, is responsible of the reduction in output absorbance values.

In equation 2, the term '$A_0$' represent a constant that depends on the specific nucleoprotein, the tissue in which its located and the measurement method that was used.

Naturally, the power W applied to the tissue during electrical treatment must be a function of different forms of energy into which the electrical energy is transformed: rotational energy (Er) and vibrational (Ev) of molecules, dissipated heat function of RT, etc, that is:

$W = f(RT) + g_1(E_v) + g_2(E_r) + ... \approx f(RT) = m\,RT$, where $m$ is a constant of proportionality expressed in $s^{-1}$.

in our case, the measurement of the current was made 5 minutes after the beginning of the treatment. For this moment the value of electrical resistance of the nodal segments is smaller than the one they had at the beginning (we can verify it measuring the electrical current at the beginning and at the end of the treatment), reason why their behavior resembles more to the one of an electrical conductor. Since the electrical power is for the most part transformed in heat, that is absorbed by the tissue, the equation 1 is not remarkably altered if we replaced RT by W. This leads to the practical expression:

$$A = A_0^{'} \frac{1}{W} \quad (2)$$

Equations 1 and 2 indicate that the greater impact in the denaturation (inactivation) of the virus is caused by the increase of the temperature inside the cell, it means that the effect of viral diseases cleaning that produces the electrical current in the vegetal tissue is, fundamentally, an effect of thermotherapy by the temperature that the viral nucleoprotein in the thermal bath is put under. This is not the chamber temperature to which the tissue is subjected, as it happens in the design of the traditional thermotherapy experiments. Each tissue, according to its physiological conditions, such as the age, volume, hydric potential, etc, will be at different temperatures. Nevertheless, these equations indicate that variable W is a good estimator of the dissipated heat by moles of particles. In this sense, the electrotherapy method is superior to the traditional thermotherapy methods.

The equations analyzed throughout the presented reasonings suggest in addition the existence of an "effective" temperature (inside the tissue) from which the probability of the occurrence of denaturation events of viral particles is high, happening then the cleaning event.

The denaturation effect can happen to all proteins in the vegetal tissue, not only to viral nucleoproteins, and evidently it can occur in a reversible fashion as soon as the application of the electrical voltage finishes to the vegetal tissue. The occurrence of the cleaning effect happens only when the viral nucleoproteins are affected and not other ones necessary for the cellular metabolism. Each protein must have a specific 'effective' temperature. Evidently, the grater the molecular weight the smaller the energy (temperature) needed to cause inactivation. The nucleoproteins that are within the cell are thermally 'isolated' by the cell wall and must be less probable the occurrence of inactivation in this site. Then, the viral particles denaturation must probable occur during their traffic for the apoplastic space, movement responsible of the desease spread. The cleaning of this intercellular space, blockage the further penetration to healthy cells, a process that is mediated by the recognition of specific nucleoprotein three-dimensional motifs that now are absent in the heat inactivated viral particles. The former support the practical evidence that when meristematic tissue (with high cellular division rate which is not yet reached by the disease) from treated explants is further cultivated, the regenerated *in vitro* plants are healthy.

**Lineal regression of experimental data. Fit to theoretical model.**

The output regression parameters if we assume a model equal to: (Y=bx+c+error, where x=1/W) are shown in tables 2 and 3.

Tables 2 and 3. Output regression parameters for the model with intercept.

| R | Square R | Adjusted Square R | Estimated Std. Error | Durbin-Watson |
|---|---|---|---|---|
| .932* | .868 | .861 | .06371 | 1.729 |

*Predictors (constant) = 1/W
Note: dependent variable = absorbance

| Model | Non standardized Coefficients | | Standardized Coefficients | t | Significance | Confidence Interval | |
|---|---|---|---|---|---|---|---|
| | B | Standard Error | Beta | | | Lower | Higher |
| Constant | -.010 | .024 | | -.417 | .681 | -.059 | .040 |
| 1/W | .141 | .013 | .932 | 11.185 | .000 | .114 | .167 |

In the case of an assumed model equal to: (Y=bx+ error, where x=1/W) the corresponding parameters are shown in tables 4 and 5.

Tables 4 and 5. Output regression parameters for the model without intercept.

| R | Square R | Adjusted Square R | Estimated Std. Error | Durbin-Watson |
|---|---|---|---|---|
| .973* | .947 | .944 | .06238 | 1.665 |

*Predictors (constant) = 1/W
Note: dependent variable = absorbance

| Model | Non standardized Coefficients | | Standardized Coefficients | t | Significance | Confidence Interval | |
|---|---|---|---|---|---|---|---|
| | B | Standard Error | Beta | | | Lower | Higher |
| 1/W | .137 | .007 | .973 | 18.828 | .000 | .121 | .152 |

Tables 2 y 4 show valid regression indexes for both models, although the value of Square R is superior, numerically, for the model Y=bx+error. However in table 3 can be observed that the constant c of the model with intercept is not significant. Then we can conclude that the statistical correct relation between ¨y¨ (the output absorbance value) and ¨x¨ (the inverse of the applied electrical power) is in the way y=bx. This relation coincides with the results of the theoretical analysis previously developed. The similarity of results among the theoretical and statistical models is a confirmation of the reliability of the experimental data obtained in this experiment.

The hyperbolic function obtained (A=0.137/P) allow to estimate the optimum voltage value in order to obtain the higher efficiency in the viral disease cleaned program (González 2005). For the improvement of this estimation is also important the homogeneity of the treated explants, because variations in their electrical resistance makes some individuals not to follow the populations statistical behavior.

**CONCLUSIONS**

We observed that an electrotherapy experiment could be reduced to a temperature controlled thermotherapy at the cellular level starting from which the denaturalization mediated inactivation of the viral nucleoprotein happens.  We suggest the viral particles inactivation preferentially occurs during their traffic for the apoplastic space instead at their stay inside the cell. The hyperbolic dependence obtained between the Absorbance Value and the Electrical Power applied to each explant (A=0.137/P) can be used in the refinement of electrotherapy mediated viral disease cleaning  experiments to minimize the intrinsic sources of  heterogeneity among the vegetal treated material.